\documentclass[aps,prl,showpacs,notitlepage,singlecolumn,superscriptaddress,11pt,floatfix,]{revtex4-1}
\usepackage{amsmath}
\usepackage{amssymb}
\usepackage{amsfonts}
\usepackage{stmaryrd}
\usepackage{wasysym}
\usepackage{graphicx}
\usepackage{multirow}
\usepackage{tabularx}
\usepackage{textcomp}
\usepackage{subfigure}
\usepackage{url}
\usepackage[colorlinks=true,citecolor=blue,urlcolor=blue]{hyperref}
\usepackage{romannum}
\usepackage{mathtools}
\usepackage[utf8]{inputenc}
\usepackage{braket}
\usepackage{amsmath}
\usepackage{xcolor}
\usepackage{colortbl}
\usepackage{color,soul} 
\usepackage{bm}
\usepackage[normalem]{ulem}
\usepackage[utf8]{inputenc}
\usepackage{array}
\usepackage{makecell}
\usepackage{multirow}
\usepackage{enumitem}  
\hypersetup{colorlinks=true, urlcolor=blue, citecolor=cyan, pdfborder={0 0 0}}

\usepackage{fancyhdr}
\usepackage{pdfpages} 
\usepackage{pgffor} 

\makeatletter
\AtBeginDocument{\let\LS@rot\@undefined}
\makeatother
\def\supplementfilename{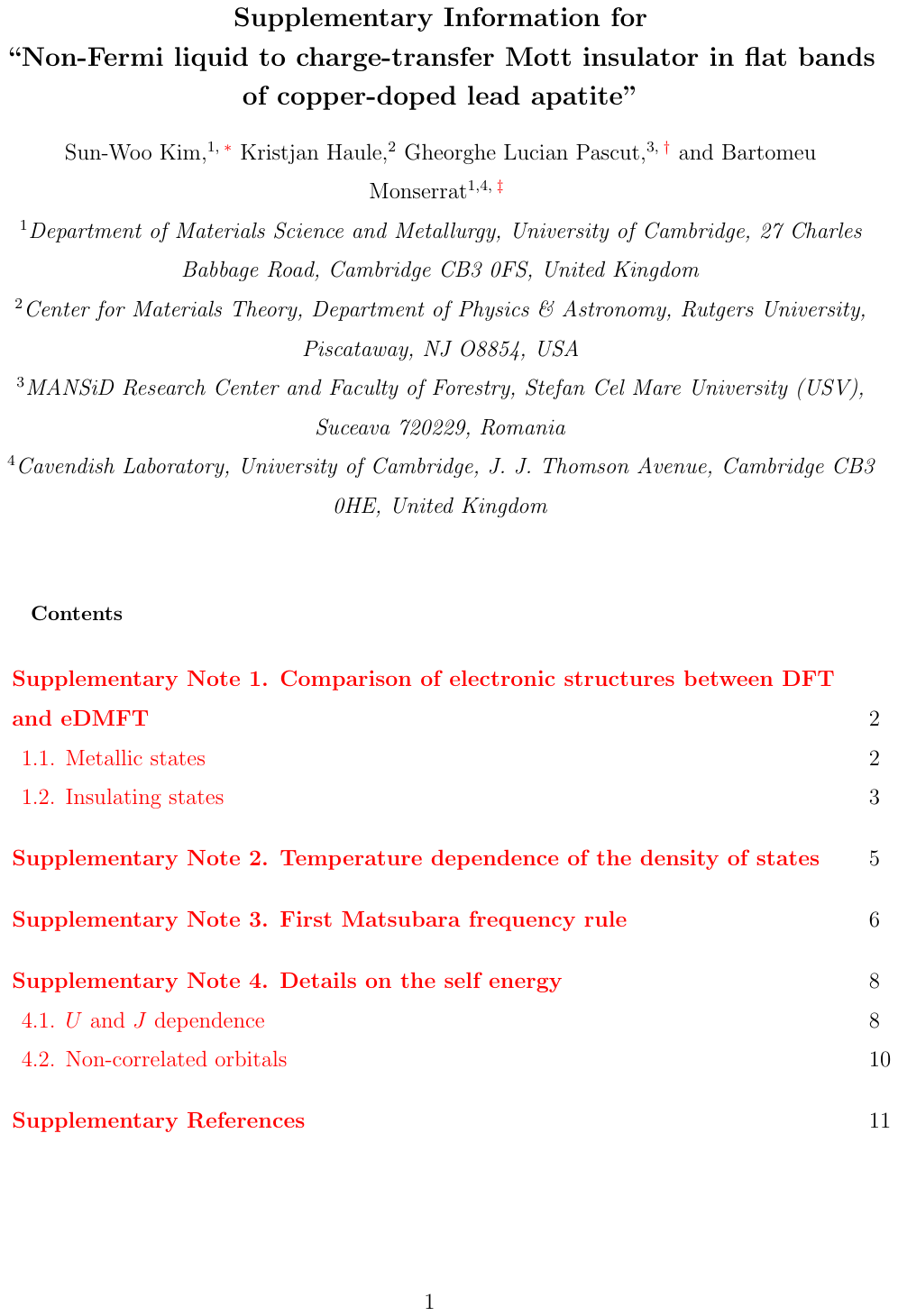}

\pdfximage{\supplementfilename}
\def\numbersupplementpages{\the\pdflastximagepages}

\newif\ifarXiv
\arXivtrue 

\begin{document}
\pagenumbering{arabic}

\title{Non-Fermi liquid to charge-transfer Mott insulator in flat bands of copper-doped lead apatite}
\maketitle
\begin{center}

{Sun-Woo Kim,$^{1,\;\textcolor{red}{*}}$ Kristjan Haule,$^{2}$ Gheorghe Lucian Pascut,$^{3,\;\textcolor{red}{\dagger}}$ and Bartomeu Monserrat$^{1,4,\;\textcolor{red}{\ddagger}}$
}

\emph{$^{1}$Department of Materials Science and Metallurgy, University of Cambridge,\\ 27 Charles Babbage Road, Cambridge CB3 0FS, United Kingdom}\\
\emph{$^{2}$Center for Materials Theory, Department of Physics \& Astronomy,\\ Rutgers University, Piscataway, NJ
O8854, USA}\\
\emph{$^{3}$MANSiD Research Center and Faculty of Forestry,\\ Stefan Cel Mare University (USV),
Suceava 720229, Romania}\\
\emph{$^{4}$Cavendish Laboratory, University of Cambridge,\\ J. J. Thomson Avenue, Cambridge CB3 0HE, United Kingdom}\\

$^{\textcolor{red}{*}}$ \textup{\href{mailto:swk38@cam.ac.uk}{swk38@cam.ac.uk}}
$^{\textcolor{red}{\dagger}}$ \textup{\href{mailto:glucian.pascut@usm.ro}{glucian.pascut@usm.ro}}
$^{\textcolor{red}{\ddagger}}$ \textup{\href{mailto:bm418@cam.ac.uk}{bm418@cam.ac.uk}}
\end{center}

\section{Abstract}

Copper-doped lead apatite, called LK-99, was initially claimed to be a room temperature superconductor driven by flat electron bands, but was later found to be a wide gap insulator.  
Despite the lack of room temperature superconductivity, there is growing evidence that LK-99 and related compounds host various strong electron correlation phenomena arising from their flat electron bands.
Depending on the copper doping site and crystal structure, LK-99 can exhibit two distinct flat bands crossing the Fermi level in the non-interacting limit: either a single or two entangled flat bands.
We explore potential correlated metallic and insulating phases in the flat bands of LK-99 compounds by constructing their correlation phase diagrams, and find both non-Fermi liquid and Mott insulating states.
We demonstrate that LK-99 is a charge-transfer Mott insulator driven by strong electron correlations, regardless of the flat band type.
We also find that the non-Fermi liquid state in the multi-flat band system exhibits strange metal behaviour, while the corresponding state in the single flat band system exhibits pseudogap behaviour.
Our findings align with available experimental observations and provide crucial insights into the correlation phenomenology of LK-99 and related compounds that could arise independently of superconductivity.
Overall, our research highlights that LK-99 and related compounds offer a compelling platform for investigating correlation physics in flat band systems.



\section{Introduction}

An electronic flat band refers to the energy dispersion of electrons that is confined within a narrow energy window in momentum space, originating from strong localization of electrons in real space.
The flat band features a high density of states and quenched kinetic energy, which amplifies electron correlation effects.
This makes flat band systems an interesting platform for studying exotic many-body quantum phenomena including the fractional quantum Hall effect\,\cite{FQHE-1-observation,FQHE-FCI}, ferromagnetism\,\cite{Flat_band_FM}, excitonic insulators\,\cite{Flat_band_exciton}, Mott insulators\,\cite{Mott_1,Mott_2}, non-Fermi liquid behaviour\,\cite{Flat_band_NFL_1}, and unconventional (potentially high-temperature) superconductivity\,\cite{Flat_band_SC_1,Flat_band_SC_2}.
Archetypal examples of flat band materials include Landau levels in two-dimensional materials under a magnetic field\,\cite{Landau1930,Klitzing-LL} and heavy fermion systems with localized $f$-electrons\,\cite{Heavy_fermion_1}.
Recent advances in twisted moiré systems\,\cite{TBG1-SC,TBG2-Mott,Balents2020,Checkelsky2024} and kagome materials\,\cite{kagome-CoSn,kagome-Ni3In} with high tunability have opened a new chapter in the field of flat band materials, stimulating researchers to explore the strongly correlated physics in these materials with the goal of understanding the mechanisms of high-temperature unconventional superconductors and moving toward the dream of room-temperature superconductivity.

The copper-doped lead apatite compound, known as LK-99, was initially claimed to be a room-temperature superconductor\,\cite{LK99-1,LK99-2}. Early theoretical studies based on density functional theory (DFT) calculations for a simple model of the LK-99 compound fueled interest by identifying very narrow flat bands crossing the Fermi level, suggesting possible high-temperature flat band superconductivity\,\cite{griffin2023origin,si2023electronic,LAI202466,kurleto2023pb}. However, subsequent experimental studies showed that LK-99 is an insulator rather than a superconductor\,\cite{Liu_2023_semiconducting,Guo2023,kumar2023absence,Liu_2023_phases,puphal2023single,Wang2023,zhang2023ferromagnetism}, and clarified that the observed resistivity drop originally associated with a superconducting transition was instead caused by a first order structural phase transition of Cu$_2$S impurities\,\cite{Jain2023,Cu2S_matter}.
The insulating nature of LK-99 is also supported by DFT calculations including ferromagnetism\,\cite{bai2023ferromagnetic,swift2023comment,pashov2023multiple} and dynamical mean-field theory calculations without long-range magnetic order\,\cite{korotin2023electronic,si2308pb10,yue2023correlated}.
Various theoretical estimates of critical superconducting temperatures in (unrealistic) metallic models yielded values significantly lower than room temperature\,\cite{oh2023s,witt2023no,paudyal2024implications}.

Despite the lack of high-temperature superconductivity, there is growing evidence that the LK-99 compound represents an interesting class of flat band materials for exploring strongly correlated physics.
In a non-interacting limit, it exhibits ideal conditions to explore correlation effects from flat bands, as these flat bands are located exactly at the Fermi level with partial filling and are isolated from other bands, similar to those in twisted moire systems.
What makes the copper-doped lead apatite system particularly interesting is that it hosts two distinct flat bands, either two entangled flat bands or a single flat band crossing the Fermi level.
The two flat bands appear in the originally claimed LK-99 structure, while the single flat band appears either at potentially low temperatures or in a different structure with copper doping at a different site\,\cite{Kim2024}.
Experimental observations of insulating states without long-range magnetic order in the LK-99 samples\,\cite{Liu_2023_semiconducting,Guo2023,kumar2023absence,Liu_2023_phases,puphal2023single} indeed support the correlated insulating states from flat bands.
Moreover, recent experimental observations of correlated metallic phases with non-Fermi liquid (NFL) behaviour in sulfur-doped LK-99 samples\,\cite{LK_strange_metal} 
further substantiate the interesting strong correlation physics.


In this work, we study correlated metallic and insulating phases in two distinct flat band systems in copper-doped lead apatite and construct their correlation phase diagrams. 
We demonstrate the LK-99 compound is a charge-transfer Mott insulator with a wide gap due to strong electron correlations in both flat band systems.
Additionally, we reveal how the correlated Mott insulating state evolves from correlated metallic states, specifically NFL states, as the electron correlation strength increases.
Furthermore, we discover that the NFL states have drastically different characteristics in the two flat band systems.
Specifically, the NFL state in the two flat bands system exhibits strange metal behaviour while the NFL state in the single flat band system exhibits pseudogap behaviour.
Our findings of both correlated metallic and insulating states are consistent with available experimental observations for both LK-99 and LK-99 related compounds. 
Our work provides crucial insights into the correlation phenomenology in non-superconducting LK-99 compounds, highlighting LK-99 as an interesting flat band material for studying strongly correlated physics arising from flat bands.

\section{Correlation phase diagram of copper doped lead apatite}

\begin{figure}[t]
 \centering
 \includegraphics[width=\textwidth]{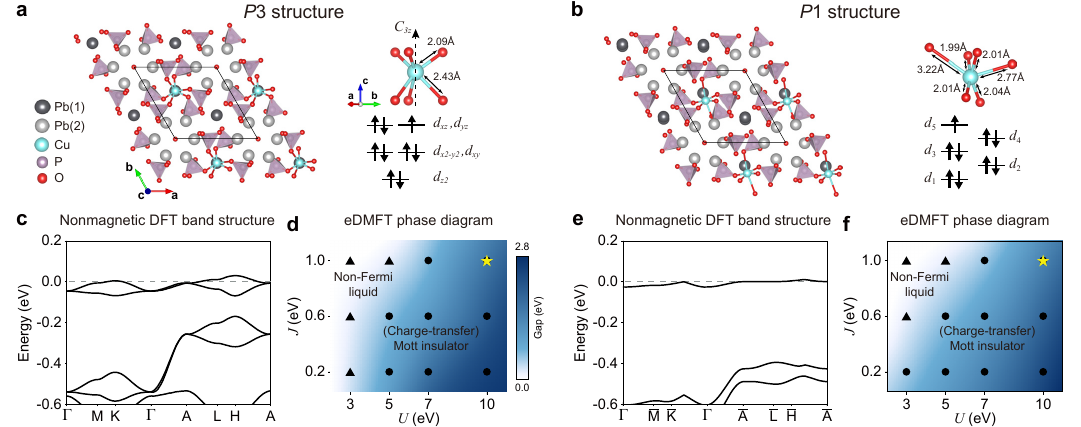} 
  \caption{ \textbf{Crystal structures, electronic structures, and correlation phase diagrams of copper-doped lead apatite.}
  \textbf{a,b} Crystal structures of Pb$_{9}$Cu(PO$_4$)$_6$O with \textbf{a} $P3$ space group and \textbf{b} $P1$ space group. 
  \textbf{c} DFT band structure for the $P3$ structure.
\textbf{d} eDMFT phase diagram in the Hubbard $U$ and Hund's coupling $J$ parameter spaces for the $P3$ structure.
  \textbf{e} DFT band structure for the $P1$ structure.
  \textbf{f} eDMFT phase diagram in the $U$ and $J$ parameter spaces for the $P1$ structure.
  In \textbf{d,f}, the color bar indicates the Mott gap near the Fermi level. 
  Triangles represent metallic states, while circles represent insulating states.
  The star indicates the constrained eDMFT values of $U$ and $J$. 
  }
  \label{fig:structure_phase_diagram}%
\end{figure}

The parent lead apatite compound, Pb$_{10}$(PO$_4$)$_6$O, crystallizes in a hexagonal lattice with space group $P3$.  
There are two symmetrically distinct lead sites, labelled Pb(1) and Pb(2) (see Fig.\,\ref{fig:structure_phase_diagram}\textbf{a}), and previous experiments observed that copper can be doped at both sites\,\cite{puphal2023single,Bernevig_phonon}. In this work, we focus on copper doping at the Pb(1) site only, motivated by claims associating this particular doping site with room-temperature superconductivity\,\cite{LK99-1,LK99-2}, and also by its intriguing electronic and structural properties, which include two entangled narrow energy bands crossing the Fermi level at the DFT level without a magnetic order and a possible structural phase transition at low temperatures to a low-symmetry structure, as predicted by theory\,\cite{Kim2024,cabezas2023theoretical}. 
It is also worth noting that the electronic structure of the low-symmetry phase with the Pb(1) doping is similar to the electronic structure of the Pb(2) doped phase\,\cite{Kim2024}, suggesting that the study of the Pb(1) case allows us to encompass a broad range of correlation phenomenology in the copper-doped lead apatite system.

Copper doping at the Pb(1) site results in a structure that preserves the $P3$ space group of the parent compound (Fig.\,\ref{fig:structure_phase_diagram}\textbf{a}). A previous study demonstrated that this structure is stable at room temperature due to anharmonic phonon-phonon interactions, and that at lower temperature it may undergo a structural phase transition to a low-symmetry structure with the $P1$ space group\,\cite{Kim2024} (Fig.\,\ref{fig:structure_phase_diagram}\textbf{b}).
In the $P3$ structure, the oxygen octahedra surrounding the copper atom exhibit  C$_{3z}$ symmetry, resulting in two different Cu-O bond lengths of $2.09\,\text{\AA}$ and $2.43$\,$\text{\AA}$.  In contrast, in the $P1$ structure, this symmetry is broken, leading to significantly different Cu-O bond lengths ranging from $1.99\,\text{\AA}$ to $3.22$\,$\text{\AA}$.

In the local atomic picture, the oxygen octahedral crystal field in the $P3$ structure leads to three different energy subspaces for the copper $d$ orbitals: a $d_{z^2}$ orbital, two degenerate $d_{x^2-y^2}$ and $d_{xy}$ orbitals, and two degenerate $d_{xz}$, and $d_{yz}$ orbitals; arranged in ascending order of energy (Fig.\,\ref{fig:structure_phase_diagram}\textbf{a}).
In the $P1$ structure, the degeneracy of copper $d$ orbitals is completely lifted and we label the five $d$ orbitals as $d_1$, $d_2$, ..., and $d_5$ in ascending order of energy (Fig.\,\ref{fig:structure_phase_diagram}\textbf{b}).
The copper atom has a $d^9$ configuration, and electron counting leads to fully occupied $d_{z^2}$, $d_{x^2-y^2}$, and $d_{xy}$ orbitals and a to a partial occupation of the degenerate $d_{xz}$ and $d_{yz}$ orbitals in the $P3$ structure. 
For the $P1$ structure, electron counting leads to four $d$ orbitals being fully filled, and the higher energy $d_5$ orbital being only half-filled.


The local atomic picture is robust, resulting in DFT electronic structures that exhibit narrow flat bands near the Fermi level. This is due to the large distance between copper atoms in both structures (at least 7.4\,\AA), leading to well-localized copper $d$ orbitals.
These atomically localized flat bands are similar to those found in $4f$ states in heavy fermion systems\,\cite{Heavy_fermion_1}, and are distinct from flat bands caused by destructive interference effects in specific lattice structures such as Lieb and kagome lattices\,\cite{Checkelsky2024}.
The electronic band structure of the $P3$ structure hosts \textit{two flat} bands with 3/2 filling crossing the Fermi level (Fig.\,\ref{fig:structure_phase_diagram}\textbf{c}), while that of the $P1$ structure features a half-filled \textit{single flat} band at the Fermi level (Fig.\,\ref{fig:structure_phase_diagram}\textbf{e}).
We emphasize that the above description, based on simple electron counting and DFT band structures, does not rigorously consider electron correlations.
Below, we discuss the importance of the Hubbard $U$ and Hund's coupling $J$ in determining the actual occupancy of the copper $d$ orbitals and the resulting correlated electronic structures in copper-doped lead apatite.

We explore the role of electron correlations on the electronic structure of copper-doped lead apatite by investigating the DFT+embedded dynamical mean-field theory (eDMFT) phase diagrams of the two structures in the space spanned by the Hubbard $U$ and Hund's coupling $J$ parameters (see Figs.\,\ref{fig:structure_phase_diagram}\textbf{d},\textbf{f}). The phase diagrams of the two structures exhibit striking similarities. At small $U$ and $J$ values both structures are metallic, with a slightly greater metallic region observed for the $P3$ structure. As the Hubbard $U$ increases, both systems become insulators, with their gaps increasing accordingly. Increasing the Hund's coupling $J$ has the opposite effect to that of Hubbard $U$ by enhancing metallicity. Although, in general, $J$ compensates for the effect of $U$, it is rather unexpected that small changes in $J$ values significantly affect the electronic properties, driving a metal-to-insulator transition as shown in the phase diagrams. 
This is notable because Hund's coupling $J$ is typically crucial in multiorbital (or $d^8$) systems, rather than in this one-hole $d^9$ system like the cuprate superconductor\,\cite{Hund_haule,Hund_medici,Hund_Georges}.
The insulating states are paramagnetic (charge-transfer) Mott insulators while the metallic states are non-Fermi liquids, which will be clarified below.

\begin{figure}[b]
 \centering
 \includegraphics[width=\textwidth]{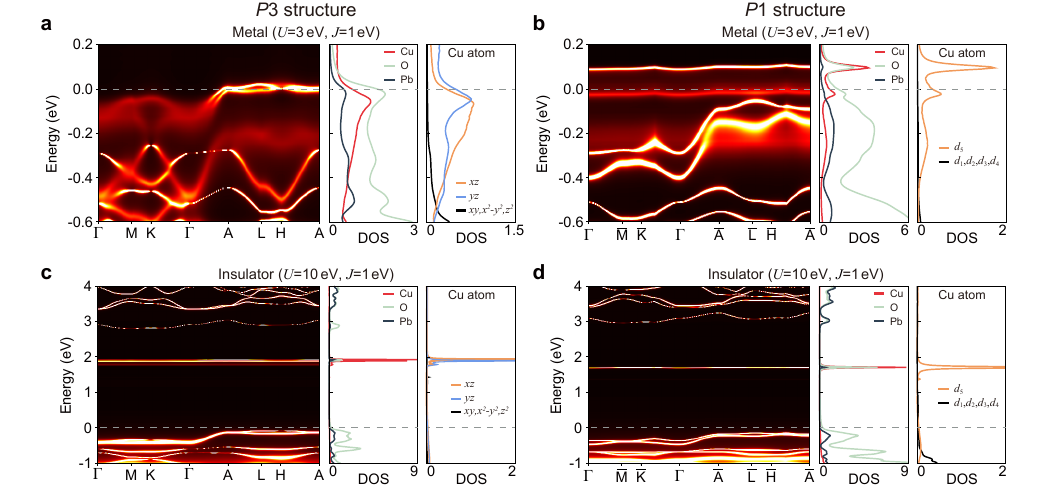} 
  \caption{
  \textbf{Metallic and insulating electronic structures of copper-doped lead apatite.}
  \textbf{a,b} Spectral function and density of states (DOS) for metallic states in the \textbf{a} $P3$ structure and \textbf{b} $P1$ structure. $U=3$\,eV and $J=1$\,eV are used.
  \textbf{c,d} Spectral function and DOS for insulating states in the \textbf{c} $P3$ structure and \textbf{d} $P1$ structure. $U=10$\,eV and $J=1$\,eV are used.
  All data are obtained at a temperature of 300\,K.
  }
 \label{fig:representative_electronic_Structures}%
\end{figure}

The \textit{ab initio} constrained eDMFT values of $U$ and $J$ are estimated as $U=10$\,eV and $J=1$\,eV. At these values for $U$ and $J$, our theory predicts that both structures are wide gap Mott insulators with a gap of 1.8\,eV for the $P3$ structure and of 1.6\,eV for the $P1$ structure (star in the phase diagrams in Figs.\,\ref{fig:structure_phase_diagram}\textbf{d} and \ref{fig:structure_phase_diagram}\textbf{f}, and see also Fig.\,\ref{fig:representative_electronic_Structures} for detailed electronic structures).
Our results are consistent with experimental reports showing highly insulating behaviour\,\cite{Liu_2023_semiconducting,Guo2023,kumar2023absence,Liu_2023_phases,puphal2023single,Wang2023,zhang2023ferromagnetism}, as well as with previous theoretical studies that also included electron correlations effects\,\cite{korotin2023electronic,si2308pb10,yue2023correlated}.
In contrast to previous DFT results\,\cite{bai2023ferromagnetic,swift2023comment,pashov2023multiple}, we do not need to assume a ferromagnetic state to obtain an insulating phase.
Moreover, compared to previous Wannier-based DMFT studies\,\cite{korotin2023electronic,si2308pb10,yue2023correlated}, our predicted Mott gap is larger. 

In what follows, we explore in detail both metallic and insulating states. We propose copper-doped lead apatite as a prototypical example of a wide gap insulator with transition metal atom doping, where dopants form localized impurity-like states, exhibiting narrow flat bands.
We anticipate that different dopants or external perturbations, such as hydrostatic pressure or strain, which can be used to suppress electron correlations, would enable the exploration of our proposed $U$-$J$ phase diagram.
Later, we will indeed discuss the accessibility of metallic states depicted in our phase diagram, as examined in recent experiments involving sulfur-doped LK-99 samples \cite{LK_strange_metal}.

We present representative correlated electronic structures of metallic and insulating states for the two structures at room temperature in Fig.\,\ref{fig:representative_electronic_Structures}.  
For metallic states ($U=3$ and $J=1$\,eV; see Figs.\,\ref{fig:representative_electronic_Structures}\textbf{a,b}), the electronic structures of both structures show significant spectral weight near the Fermi level, primarily composed of copper and oxygen orbitals. 
The density of states (DOS) for the $P3$ structure exhibits a single sharp peak, whereas the DOS for the $P1$ structure displays two peaks.
The relevant copper $d$ orbitals for these states are the $d_{xz}$ and $d_{yz}$ orbitals for the $P3$ structure and the $d_5$ orbital for the $P1$ structure, consistent with the local atomic picture and DFT band structures discussed earlier.
However, the eDMFT electronic structures of the metallic phases are markedly different from those obtained from DFT in Figs.\,\ref{fig:structure_phase_diagram}\textbf{c,e} (see also Supplementary Note 1 for a broader energy window comparison). As discussed below in detail, this difference is caused by the fact that these are highly correlated metallic states exhibiting non-Fermi liquid behaviour, as hinted by the incoherent nature of the spectral functions around the Fermi level.
We also note that the spectral functions of the two structures are distinct near the Fermi level. The $P3$ structure exhibits a mixture of incoherent and coherent spectral features depending on the Brillouin zone (BZ) region, with particularly coherent states at the Fermi level. By contrast, the $P1$ structure exhibits highly incoherent states at the Fermi level across the whole BZ, showing a pseudo-gap like electronic structure.

For insulating states ($U=10$ and $J=1$\,eV; see Figs.\,\ref{fig:representative_electronic_Structures}\textbf{c,d}), the copper orbitals undergo significant changes compared to the metallic states, forming a charge-transfer Mott gap of 1.8 and 1.6\,eV for the $P3$ and $P1$ structures, respectively. The relevant copper states comprise the flat lowest unoccupied states, characterized by a very narrow bandwidth. These unoccupied flat bands are located within the fundamental bulk gap of the parent lead aptite compound (without copper doping) formed by lead and oxygen states, with a value of 2.7\,eV for the $P3$ structure and of 2.9\,eV for the $P1$ structure.
The bulk gap of the $P3$ structure is almost identical to that of the parent lead apatite bulk gap, and this gap becomes larger in the $P1$ structure due to the distortion of lead and oxygen atoms.
The formation and evolution of the copper flat bands are detailed below.
It is noteworthy that eDMFT electronic structures of the insulating phases are also different from DFT (see Supplementary Note 1), as the eDMFT calculations are performed for paramagnetic states, rather than the long-range ferromagnetic states required in DFT-level calculations to open a gap.

\begin{figure}[t]
 \centering
 \includegraphics[width=.6\textwidth]{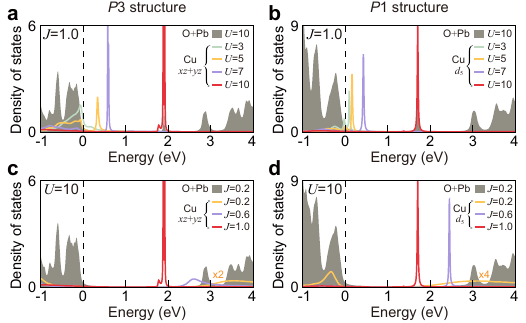} 
  \caption{
  \textbf{Influences of Hubbard $U$ and Hund's coupling $J$ on copper $d$ orbital states.} \textbf{a,b} DOS of copper $d$ orbitals for the \textbf{a} $P3$ structure and \textbf{b} $P1$ structure as a function of Hubbard $U$. $J=1.0$\,eV is used. \textbf{c,d} DOS of copper $d$ orbitals for the \textbf{c} $P3$ structure and \textbf{d} $P1$ structure as a function of Hund's coupling $J$. $U=10$\,eV is used. For comparison, the DOS of oxygen and lead states are shown in shaded grey.
  }
 \label{fig:evolution_of_dos}%
\end{figure}

Figure\,\ref{fig:evolution_of_dos} shows the evolution of the DOS of copper correlated $d$ orbitals as a function of $U$ and $J$.
With fixed $J=1$\,eV, increasing $U$ leads to the formation of a sharp narrow peak composed of $d_{xz}$ and $d_{yz}$ orbitals for the $P3$ structure as it undergoes a phase transition from a metallic to an insulating state (Fig.\,\ref{fig:evolution_of_dos}\textbf{a}). It is noteworthy that at $U=5$\,eV, the $P3$ structure remains metallic with a pseudo-gap-like DOS.
This pseudo-gap-like feature also exists in the metallic state with the peak of the $d_5$ orbital of the $P1$ structure at $U=3$ and $U=5$\,eV (with $J=1$\,eV; Fig.\,\ref{fig:evolution_of_dos}\textbf{b}).
As a result, at $U=10$\,eV (with $J=1$\,eV), both structures exhibit a sharp peak composed of the correlated $d$ orbitals within the fundamental bulk gap formed by oxygen and lead orbitals. 
This sharp peak remains robust against temperature changes from 100\,K to 1900\,K, with small shifts in its position within the fundamental gap depending on temperature (see Supplementary Note 2). 
When the $J$ decreases (so the effective $U$ increases), the sharp peak further moves towards the upper bulk states (Figs.\,\ref{fig:evolution_of_dos}\textbf{c} and \textbf{d}).
If the peak overlaps with the bulk states, it broadens and is eventually suppressed due to the hybridization with the bulk states.



\begin{figure}[t]
 \centering
 \includegraphics[width=\textwidth]{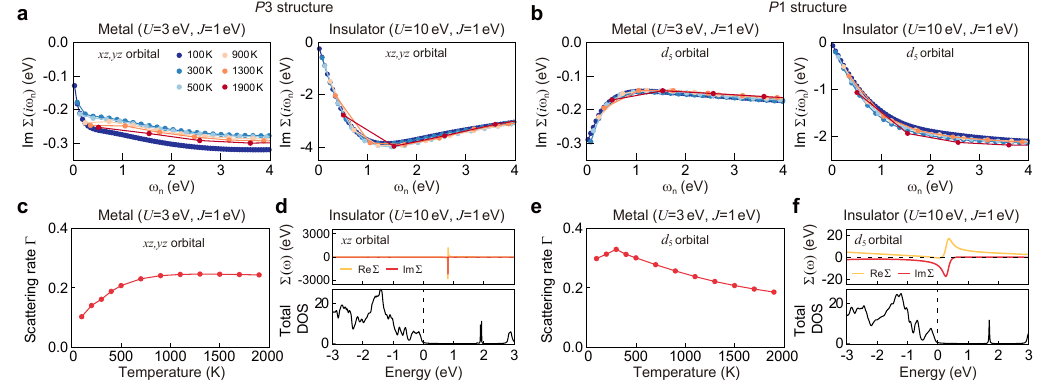} 
  \caption{
  \textbf{Characterization of non-Fermi liquid and charge-transfer Mott insulator.}
  \textbf{a,b} Temperature evolution of the imaginary part of the self energy on the imaginary axis, $\text{Im}\Sigma(i\omega_n)$, for \textbf{a} the $P3$ structure and \textbf{b} the $P1$ structure.
We display $\text{Im}\Sigma(i\omega_n)$ for the averaged $d_{xz}$ and $d_{yz}$ orbitals in the $P3$ structure, and the $d_5$ orbital in the $P1$ structure, respectively (see Supplementary Note 4 for the self-energy of other orbitals).
  \textbf{c} Scattering rate of the metallic state of the $P3$ structure as a function of temperature.
  The scattering rate is averaged over the $d_{xz}$ and $d_{yz}$ orbitals.
  \textbf{d} Self energy on the real axis and total DOS of the insulating state of the $P3$ structure.
  \textbf{e} Scattering rate of the metallic state of the $P1$ structure as a function of temperature. 
 In \textbf{c,e}, the scattering rate $\Gamma \sim -\text{Im}\Sigma(i0^+)$ is obtained using a linear fit over the first two Matsubara frequencies.
   \textbf{f} Self energy on the real axis and total DOS of the insulating state of the $P1$ structure.
  }
 \label{fig:SE_metal_insulator}%
\end{figure}

We further characterize the metallic and insulating states by examining their self-energy in detail (Fig.\,\ref{fig:SE_metal_insulator}). Our analysis reveals that both structures exhibit non-Fermi liquid (NFL) behaviour in their metallic states, with distinct characteristics between the two structures hinted at above in the discussion of their electronic structures in Fig.\,\ref{fig:representative_electronic_Structures}.
The NFL nature is identified by the low-frequency behaviour of the imaginary part of the local Matsubara self-energy, $\text{Im}\Sigma(i\omega_n)$: it deviates from linear behaviour at low temperature as $\text{Im}\Sigma(i\omega_n)\sim A(i\omega_n)^\alpha$ with $\alpha\neq1$ and its intercept $\text{Im}\Sigma({i\omega_n \to 0^+})$ shows a non-quadratic temperature dependence. 
We also confirm the NFL nature using the first Matsubara frequency rule (see Supplementary Note 3).
The intercept and the slope $A$ represent the scattering rate and the quantity directly associated with quasiparticle mass renormalization, respectively, in Landau Fermi liquid theory. Although the quasiparticle concept is generally lost in NFLs, it remains meaningful in certain types of NFLs with a negative slope $A<0$ of $\text{Im}\Sigma(i\omega_n)$, where the negative slope is used as a measure of mass enhancement due to electron correlations. In our system, this is the case for the $P3$ structure (left panel in Fig.\,\ref{fig:SE_metal_insulator}\textbf{a}).
We refer to the NFL state in the $P3$ structure as a bad-metallic NFL, since it still shows metallic behaviour of the scattering rate, i.e., increasing as a function of $T$, notably with a linear dependence (strange metal behaviour) below 500\,K, and saturated at higher temperature (Fig.\,\ref{fig:SE_metal_insulator}\textbf{c}). 
On the other hand, the NFL state of the $P1$ structure exhibits markedly contrasting behaviour, featuring a positive slope $A>0$ of $\text{Im}\Sigma(i\omega_n)$ with negligible temperature dependence (left panel in Fig.\,\ref{fig:SE_metal_insulator}\textbf{b}).
The positive slope indicates a highly incoherent state devoid of the quasiparticle concept.
The scattering rate of the $P1$ structure shows nonmonotonic temperature dependence, particularly a decreasing behaviour with increasing $T$ (Fig.\,\ref{fig:SE_metal_insulator}\textbf{e}).
The decreasing behaviour is also reported in the NFL state in the Lieb lattice with a flat band\,\cite{kumar_2019_magnetization} and in the single band Hubbard model at half filling\,\cite{georges2004mott}.
We refer to this as the pseudogap NFL.
The pseudogap NFL in the $P1$ structure is a more correlated state than the bad-metallic NFL in the $P3$ structure as the $\text{Im}\Sigma(i\omega_n)$ of the $P3$ structure exhibits similar pseudogap NFL behaviour when $J$ is reduced (see Supplementary Note 4).
This suggests that the two entangled energy bands in the $P3$ structure result in a metallic state with reduced correlations at identical $U$ and $J$ values.


For the insulating states, both structures are characterized as charge-transfer Mott insulators.
The low-frequency behaviour of the imaginary part of the local Matsubara self-energy $\text{Im}\Sigma(i\omega_n)$ exhibits similar behaviour in both structures:  it extrapolates to zero at zero imaginary frequency, as there are no electrons to scatter, which is expected for insulating states, and it shows negligible temperature dependence (right panels in Figs.\,\ref{fig:SE_metal_insulator}\textbf{a} and \ref{fig:SE_metal_insulator}\textbf{b}).
The real and imaginary parts of the self energy on the real axis exhibit a pole within the gap for both structures (Figs.\,\ref{fig:SE_metal_insulator}\textbf{d} and \ref{fig:SE_metal_insulator}\textbf{f}), indicating an electron-correlation-driven Mott gap. Considering the atom-projected DOS of both structures (Figs.\,\ref{fig:representative_electronic_Structures}\textbf{c,d} and Fig.\,\ref{fig:evolution_of_dos}), they are characterized as charge-transfer type insulators, where the charge transfer occurs between oxygen $p$ orbitals and copper $d$ orbitals, forming the gap with valence bands composed mainly of oxygen $p$ orbitals and conduction bands composed mainly of copper $d$ orbitals. As a result, we refer to the insulating states of both structures as charge-transfer Mott insulators.

%



\begin{figure}[b]
 \centering
 \includegraphics[width=.6\textwidth]{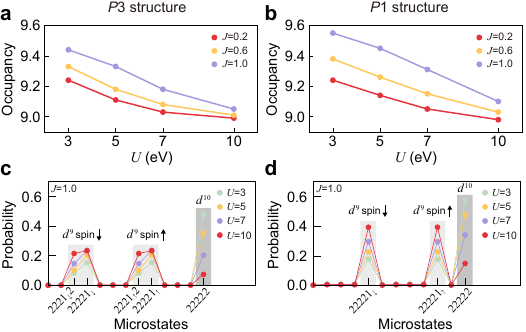} 
  \caption{
  \textbf{Occupancy of copper $d$ orbitals and their valence histogram.}  
  \textbf{a,b} Occupancy of copper $d$ orbitals as a function of $U$ and $J$ for \textbf{a} the $P3$ structure and \textbf{b} the $P1$ structure.
  \textbf{c,d} Valence histogram as a function of $U$ ($J=1$\,eV) for \textbf{c} the $P3$ structure and \textbf{d} the $P1$ structure.
    $22221_{\uparrow}$ refers to the $d^9$ electron configuration of copper, where the the first four orbitals are fully occupied (denoted as $2$), and the fifth orbital is half occupied by a spin-up electron (denoted as $1_{\uparrow}$). The orbital sequence is \{$d_{z^2}$, $d_{x^2-y^2}$, $d_{xy}$, $d_{xz}$, $d_{yz}$\} for the $P3$ structure and \{$d_{1}$, $d_{2}$, $d_{3}$, $d_{4}$, $d_{5}$\} for the $P1$ structure. 
  }
 \label{fig:occupancy_histogram}%
\end{figure}

Finally, we analyze the occupancy of copper $d$ orbitals and their microstates as the interaction parameters $U$ and $J$ change (Fig.\,\ref{fig:occupancy_histogram}).
At lower interaction parameter values ($U=3$ and $J=1$ eV), corresponding to the NFL states discussed above, the calculated copper $d$ occupancies are 9.44 and 9.55 for the $P3$ and $P1$ structures, respectively, due to a high probability of $d^{10}$ configurations.
These occupancies in NFL states deviate from a $d^9$ configuration typically assumed in general model studies based on the Cu$^{2+}$ valence state derived from the stoichiometry.
The occupancy of copper $d$ orbitals decreases as $U$ increases ($J$ decreases) for both structures, becoming 9.05 and 9.10 in the charge-transfer Mott insulator regime (at $U=10$ and $J=1$ eV) for the $P3$ and $P1$ structures, respectively, owing to the increasing probability of the $d^9$ configuration.
It is noteworthy that for the $P3$ structure, the probabilities of $d_{xz}$ and $d_{yz}$ orbitals become identical only when $U$ is sufficiently large.
This analysis provides important insights for future studies on NFL states as well as charge-transfer Mott insulating states in these prototypical flat band compounds.

\section{Discussion}




Our proposed correlation phase diagrams in two distinct flat band systems encompass a broad correlation phenomenology, ranging from correlated metallic to correlated insulating states arising from Mott physics, with direct relevance to experimental observations in LK-99 and related compounds. 
For the insulating states found in both flat band systems, our theory predicts a correlated Mott insulating phase with paramagnetism arising from local magnetic moments. This calculated insulating phase is consistent with the highly insulating nature of some LK-99 samples exhibiting paramagnetic behaviour\,\cite{Liu_2023_semiconducting,Guo2023}. 
For the newly revealed NFL states, particularly the NFL state in the two entangled flat bands in the $P3$ structure, the calculated temperature-dependent scattering rate shows a remarkable qualitative similarity to the resistivity behaviour of correlated metallic states reported in sulfur-doped LK-99 compounds\,\cite{LK_strange_metal}: $T$-linear behaviour from low temperatures (starting around 20\,K; depending on the samples) and saturation above 300\,K. 
Our theory indicates that the $T$-linear resistivity has an anomalous origin from electron correlations rather than the conventional electron-phonon coupling origin. 
We attribute the emergence of this NFL state to reduced effective electron correlations. 
We note that our theory is not compatible with the diamagnetic behaviour in some LK-99 and doped LK-99 samples, which requires further study.


The anomalous $T$-linear resistivity is a hallmark of the strange metal phase, which is known as the parent correlated metallic state of the superconducting state in high-temperature cuprates and other unconventional superconductors\,\cite{Phililp_strange}.
In this context, recent experimental work\,\cite{LK_strange_metal} argued for possible superconductivity in LK-99 and related compounds due to the observation of strange metal behaviour in sulfur-doped LK-99, analogous to other unconventional superconductors, as well as a strong diamagnetic response, reigniting the claim of the existence of superconductivity in this system\,\cite{LK_strange_metal,wang2024unveiling,wang2024indications}.
Our calculations suggest that such strange metal behaviour could arise from the strong electron correlations associated with the flat bands in the system.
However, the existence of a strange metal phase does not guarantee the presence of superconductivity, as strange metal phases have been observed in both superconducting\,\cite{strange_metal_heavy_fermion_SC,stange_metal_pnictide,strange_metal_cuprate,strange_metal_nickelate} and non-superconducting\,\cite{kagome-Ni3In,strange_metal_CeRh6Ge4} materials.
Moreover, we note that the spin exchange in the LK-99 compound is extremely small compared to that in cuprate supercondutors, as our eDMFT calculations show that the antiferromagnetic state is unstable even at 100\,K.
This small spin exchange precludes the possibility of high-temperature unconventional superconductivity similar to that observed in cuprates.
Our results contribute to settling the on-going discussion regarding the original claim of room temperature superconductivity.
Our work suggests that LK-99 compounds under doping or pressure represent additional intriguing strange metal candidates. 
Expanding the range of strange metal materials provides a valuable opportunity to explore the prerequisites for unconventional superconductivity and gain a deeper understanding of its enigmatic underlying mechanisms.

\section{Conclusions}

Our research highlights that LK-99 and related compounds provide a compelling platform for investigating correlation physics in both multi- and single-flat band systems.
Specifically, the multi-flat band system exhibits distinctly different correlated metallic phases compared to the single-flat band system, offering valuable insights into correlated phenomena within flat band materials. 
This serves as a pertinent material example amidst ongoing research into correlation effects in moiré and kagome materials, and could potentially apply to generic semiconductor systems doped with transition metal or rare-earth atoms that form highly localized states.
The correlation phase diagrams we calculate, encompassing Mott insulators and non-Fermi liquid metallic states, exhibit similarities to those observed in cuprate superconductors despite the absence of room-temperature superconductivity in LK-99 compounds. This comparison prompts future experimental and theoretical studies to explore these intriguing parallels further.


\clearpage
\section{Methods}

{\em Density functional theory (DFT) calculations. -}
We perform DFT calculations using the full-potential linearized augmented plane wave (FP-LAPW) method as implemented in the {\sc wien2k} code\,\cite{wien2k}. 
For the exchange-correlation energy, we use the generalized-gradient approximation functional of Perdew-Burke-Ernzerhof (PBE)\,\cite{PBE}. 
For Brillouin zone integrations, we use the tetrahedron integration method with 3000 and 1160 \textbf{k}-points in the full Brillouin zone, corresponding to $13\times13\times15$ and $11\times9\times10$ \textbf{k}-grids for the $P3$ structure and the $P1$ structure, respectively. We use self-consistency cycle stopping criteria of $5\times10^{-5}$\,Ry for the energy and $5\times10^{-5}$\,e for the charge. The radii (R) of the muffin-tin (MT) spheres are taken to be 1.97, 2.22, 1.41 and 1.34 Bohr for Cu, Pb, O and P atoms, respectively. R$_{\text{MT}}$$\times$K$_{\text{max}}$ is set to 5.5 (confirming that using 7 yields the same results), where K$_{\text{max}}$ is the cutoff value of the modulus of the reciprocal lattice vectors and R$_{\text{MT}}$ is the smallest MT radius.
We optimise the lattice parameters for both the $P3$ and the $P1$ structures and the internal coordinates are optimised until all forces are below $0.5$\,mRy/Bohr.

{\em DFT+embedded dynamical mean-field theory (DFT+eDMFT) calculations. -} 
We perform fully self-consistent DFT+eDMFT calculations. The iterations stop after full convergence of the charge density, the impurity level, the chemical potential, the
self-energy, and the lattice and impurity Green’s functions.
Double counting between the DFT and DMFT correlations is treated within an exact double-counting scheme\,\cite{exactDC}.
The fully rotational invariant form of the Coulomb repulsion \cite{Full_Coulomb} is employed, as it is found to be essential for accurately describing the copper-doped lead apatite compounds due to their low symmetry structures.
We use a large hybridization window of $\pm$10\,eV. For the DMFT projectors, we choose quasi-atomic localized orbitals. The radial part is determined by solving the Schrödinger equation within the muffin-tin sphere, with the linearized energy at the Fermi level, and the angular dependence provided by spherical harmonics.
For the quantum impurity problem (Cu-$3d$ electrons), we use a version of the continuous time quantum Monte Carlo (CTQMC) impurity solver\,\cite{CTQMC-1,CTQMC-2} with a total of 
$512\times$20$\times$10$^6$
Monte Carlo steps. The position of the chemical potential is allowed to vary during the self-consistent calculations. Analytic continuation from the imaginary to real frequency axis is conducted using the maximum entropy method. 

\section{Author contributions}
S.-W.K., K.H., G.L.P. and B.M. conceived the study. S.-W.K., G.L.P., and B.M. planned and supervised the research. S.-W.K. performed the DFT calculations, S.-W.K., K.H., and G.L.P. performed the eDMFT calculations. S.-W.K. and B.M. wrote the manuscript with input from all authors.

\section{Conflicts of interest}
There are no conflicts of interest to declare.

\section{Acknowledgments}

\begin{acknowledgments}
S.-W.K. thanks Tianyu Wu and Jaeyong Kim for helpful discussions.
S.-W.K. and B.M. are supported by a UKRI Future Leaders Fellowship [MR/V023926/1]. B.M. also acknowledges support from the Gianna Angelopoulos Programme for Science, Technology, and Innovation, and from the Winton Programme for the Physics of Sustainability. K.H. acknowledges funding from NSF DMR-2233892. G.L.P. acknowledges funding from the Romania National Council for Higher Education Funding, CNFIS, project number CNFIS-FDI-2024-F-0155.
The computational resources were provided by the Cambridge Tier-2 system operated by the University of Cambridge Research Computing Service and funded by EPSRC [EP/P020259/1], by the UK National Supercomputing Service ARCHER2, for which access was obtained via the UKCP consortium and funded by EPSRC [EP/X035891/1], and by the  SCARF cluster of the STFC Scientific Computing Department.

\textit{For the purpose of open access, the authors have applied a Creative Commons Attribution (CC BY) licence to any Author Accepted Manuscript version arising from this submission.}
\end{acknowledgments}

\def\bibsection{

\ifarXiv
    

\section*{Supplementary material}

# Supplementary Note 1. COMPARISON OF ELECTRONIC STRUCTURES BETWEEN DFT AND EDMFT

## 1.1. Metallic states

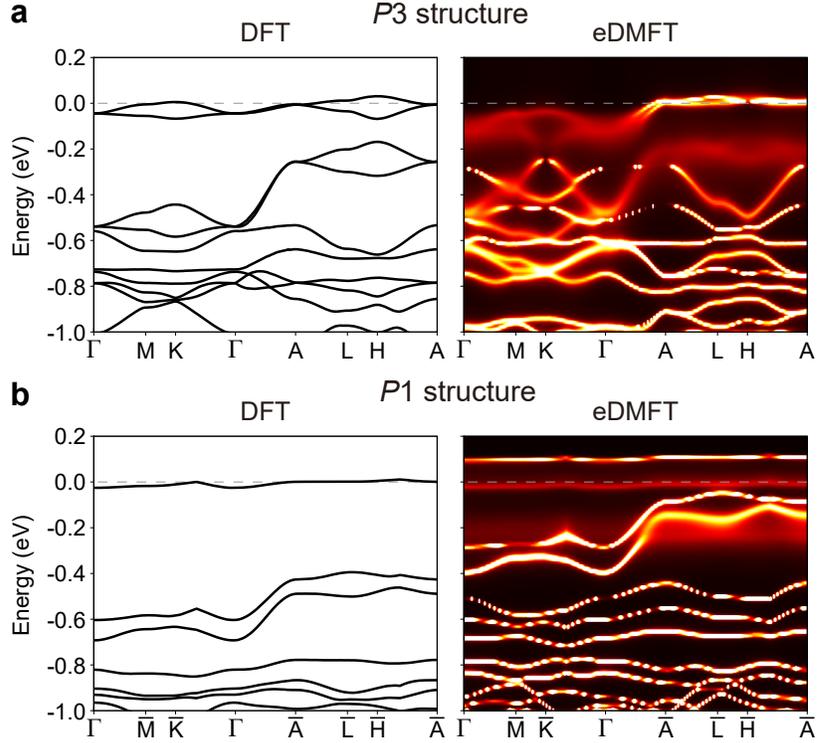

FIG. S1: **Comparison of metallic electronic structures obtained using DFT and eDMFT calculations. a,b** Nonmagnetic DFT band structure and paramagnetic eDMFT spectral function for **a** the $P3$ structure and **b** the $P1$ structure. For the DFT calculations, the PBE exchange-correlation functional is used and a nonmagnetic state is imposed. For the eDMFT calculations, $U = 3\,\text{eV}$ and $J = 1\,\text{eV}$ are used at a temperature of $300\,\text{K}$ for the paramagnetic state.

Figure S1 compares the DFT and eDMFT electronic structures for metallic states. For the $P3$ structure, both methods show two flat bands near the Fermi level (Fig. S1**a**). However, there are notable differences between the eDMFT spectral function and the single-particle DFT band structure. Specifically, the spectral function exhibits incoherent states near the Fermi level, characterized by broadened spectral weight due to electron correlations. This broadening causes certain coherent band states to appear below the Fermi level in eDMFT, which do not exist in the corresponding energy range of the DFT band structure. Additionally, eDMFT shows changes in other coherent bands with an overall bandwidth

renormalization.

For the $P1$ structure, eDMFT reduces the energy gap between the single flat band and its closest lower band, leading to the formation of highly incoherent states at the Fermi level across the entire Brillouin zone (Fig. S1**b**). These incoherent states exhibit a pseudo-gap-like density of states, which cannot be captured by DFT due to their highly correlated nature, as discussed in the main text.

As a note of clarification, we only obtain a metallic DFT band structure by imposing a non-magnetic state. The ground state corresponding to our DFT calculation is a ferromagnetic insulator (see next section). Additionally, we note that depending on the choice of $U$ in a DFT+$U$ context, we could also get a ferromagnetic metal as the ground state.

### 1.2. Insulating states

Figure S1 compares the DFT and eDMFT electronic structures for insulating states. To obtain insulating states, we introduce a ferromagnetic ordering for the DFT calculations. With our chosen Hubbard parameter of $U = 3$ eV, DFT yields Cu magnetic moments along the $z$ axis of 0.712 and 0.671 $\mu_B$ per unit cell for the $P3$ and $P1$ structures, respectively. Both methods appear to show overall agreement for both the $P3$ and $P1$ structures when only comparing the band dispersions. However, a key difference between the two methods appears when considering the spin character: in DFT, both structures show a single spin-polarized unoccupied narrow flat band and spin-polarized occupied bands, leading to an overall ferromagnetic insulating state. By contrast, eDMFT shows non-spin-polarized paramagnetic unoccupied and occupied bands.

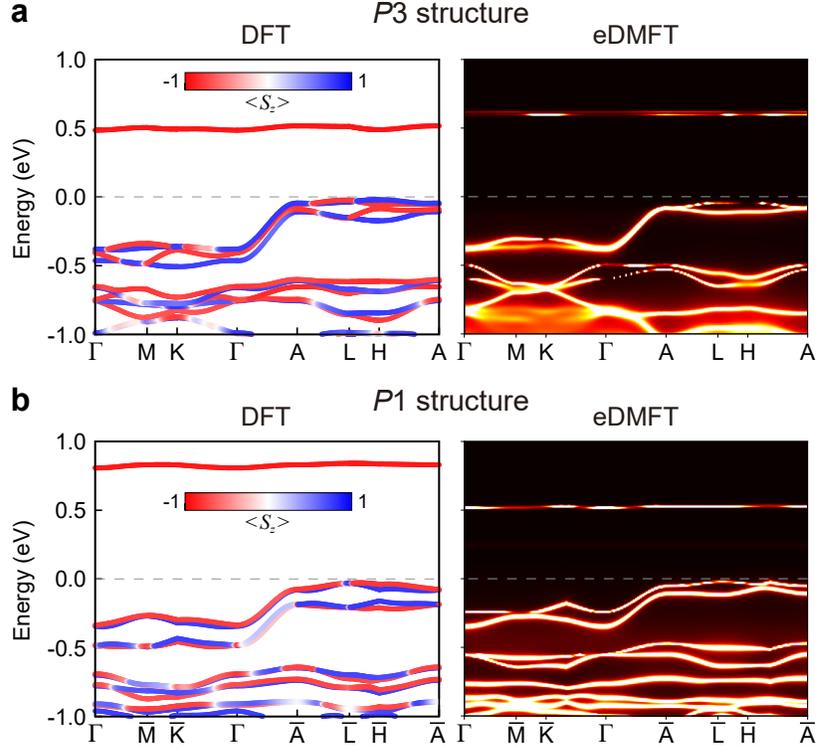

FIG. S2: **Comparison of insulating electronic structures obtained using DFT and eDMFT calculations. a,b** Ferromagnetic DFT band structure and paramagnetic eDMFT spectral function for **a** the $P3$ structure and **b** the $P1$ structure. For the DFT calculations, the PBEsol [1] exchange-correlation functional with Hubbard $U = 3\,\text{eV}$ and spin-orbit coupling is used for the ferromagnetic state, utilizing the VASP code [2, 3]. The spin-polarized DFT band structures are projected onto the spin $S_z$ component. For eDMFT calculations, $U = 7\,\text{eV}$ and $J = 1\,\text{eV}$ are used at a temperature of $300\,\text{K}$ for the paramagnetic state.

4

**Supplementary Note 2. TEMPERATURE DEPENDENCE OF THE DENSITY OF STATES**

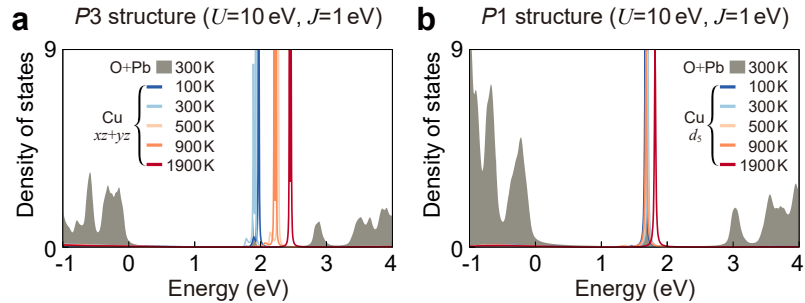

FIG. S3: **Temperature dependence of the density of states (DOS) for insulating states.** **a,b** DOS of correlated copper $d$ orbitals for the **a** $P3$ structure and **b** $P1$ structure as a function of temperature. $U = 10\,\text{eV}$ and $J = 1\,\text{eV}$ are used. For comparison, the DOS of oxygen and lead states are shown in shaded grey.

In the main text, we discuss the formation of a narrow unoccupied flat band within the fundamental bulk gap as the $P3$ or $P1$ structures undergo a metal-insulator transition with increasing $U$ and/or decreasing $J$. We demonstrate that this narrow unoccupied band in the insulating states remains positioned within the fundamental bulk gap despite wide temperature variations ranging from $100\,\text{K}$ to $1900\,\text{K}$ for both structures (Fig. S3). As temperature increases, the narrow unoccupied band blueshifts towards the conduction bulk states, with a larger shift in the $P3$ structure compared to the $P1$ structure.

**Supplementary Note 3. FIRST MATSUBARA FREQUENCY RULE**

In a Fermi liquid, the imaginary part of the self energy at the first Matsubara frequency, $\text{Im}\,\Sigma(i\omega_0)$, exhibits a $T$-linear scaling behaviour at low temperatures, which is referred to as the first Matsubara frequency rule [4]. This linear dependence arises because the self energy of a Fermi liquid at low temperatures and frequencies takes the following form on the real axis:

$$\text{Re}\,\Sigma(\omega) = (1 - Z^{-1})\omega, \quad \text{Im}\,\Sigma(\omega, T) = -C(\omega^2 + \pi^2 T^2), \tag{1}$$

where $Z$ is the quasiparticle renormalization factor and $C$ is a coefficient. Under analytical continuation to Matsubara frequencies $i\omega_n$, the imaginary part of the self energy becomes

$$\text{Im}\,\Sigma(i\omega_n, T) = (1 - Z^{-1})\omega_n + C(\omega_n^2 - \pi^2 T^2). \tag{2}$$

At the first Matsubara frequency $\omega_0 = \pi T$, the quadratic temperature term vanishes and we obtain:

$$\text{Im}\,\Sigma(i\omega_0, T) = (1 - Z^{-1})\pi T, \tag{3}$$

leaving only the linear temperature term.

Figure S4 shows the imaginary part of the self energy at the first Matsubara frequency, $\text{Im}\,\Sigma(i\omega_0)$, for metallic states in the $P3$ and $P1$ structures. For both structures, the values of $\text{Im}\,\Sigma(i\omega_0)$ for non-correlated orbitals (black) exhibit perfect linear-$T$ scaling behaviour even at higher temperatures, consistent with their Fermi liquid nature. In contrast, the values of $\text{Im}\,\Sigma(i\omega_0)$ for correlated orbitals for both structures (red), which are relevant for states near the Fermi level, deviate significantly from linearity, violating the first Matsubara frequency rule. This confirms the non-Fermi-liquid nature of the metallic states in both structures.

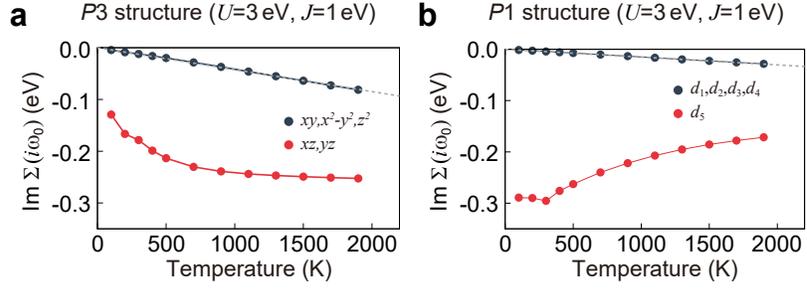

FIG. S4: **First Matsubara frequency rule. a,b** The imaginary part of the self energy at the first Matsubara frequency, $\operatorname{Im}\Sigma(i\omega_0)$, for the non-Fermi-liquid states in **a** the $P3$ structure and **b** the $P1$ structure. Red lines and symbols denote the average $\operatorname{Im}\Sigma(i\omega_0)$ for correlated orbitals, while black lines and symbols represent the corresponding average for non-correlated orbitals. Grey dotted lines indicate linear fits of $\operatorname{Im}\Sigma(i\omega_0)$ in the low-temperature regions for the non-correlated orbitals. The values of $\operatorname{Im}\Sigma(i\omega_0)$ for non-correlated orbitals exactly follow these linear fits even at higher temperatures, whereas the trends for correlated orbitals deviate significantly from linearity.

**Supplementary Note 4. DETAILS ON THE SELF ENERGY**

### 4.1. $U$ and $J$ dependence

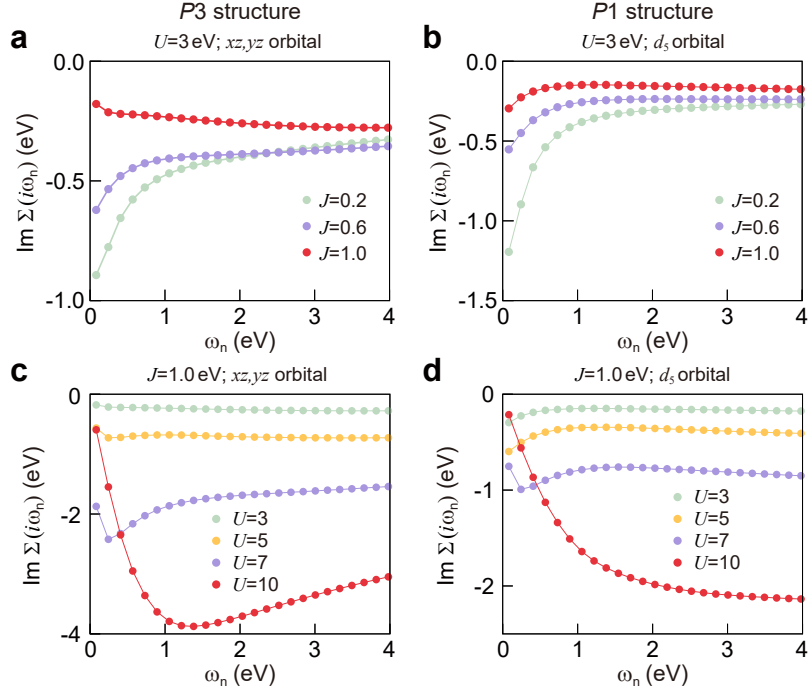

FIG. S5: **Influences of Hubbard $U$ and Hund's coupling $J$ on the self energy of correlated copper $d$ orbital states. a,b** The imaginary part of the self energy on the imaginary axis, Im$\Sigma(i\omega_n)$, at fixed $U = 3\,\text{eV}$ as a function of $J$ for **a** the $P3$ structure and **b** the $P1$ structure. **c,d** Im$\Sigma(i\omega_n)$ at fixed $J = 1\,\text{eV}$ as a function of $U$ for **c** the $P3$ structure and **d** the $P1$ structure.

In the main text, we analyze the NFL states of both structures at $U = 3\,\text{eV}$ and $J = 1\,\text{eV}$ in detail, assigning the NFL state in the $P3$ structure as a bad-metallic NFL and the NFL state in the $P1$ structure as a pseudogap NFL based on their detailed low-frequency behaviour of Im$\Sigma(i\omega_n)$, their electronic structures, and the temperature dependence of their scattering rates. A crucial factor distinguishing these two NFL states is the slope $A$ of Im$\Sigma(i\omega_n)$ at low frequencies: the bad-metallic NFL shows a negative slope $A < 0$, whereas the pseudogap NFL shows a positive slope $A > 0$.

To gain further insights into these NFL states, we investigate Im$\Sigma(i\omega_n)$ by varying $J$ values at fixed $U = 3\,\text{eV}$. As $J$ decreases, the magnitude of Im$\Sigma(i\omega_n)$ increases for both structures, reflecting enhanced electron correlations due to the increased effective $U$ (Figs. S5**a,b**). Notably, in the $P3$ structure, the slope $A$ changes sign as $J$ decreases, indicating a tran-

8

sition to a pseudogap NFL state. For $J \leq 0.6$ eV, the behaviour of $\mathrm{Im}\,\Sigma(i\omega_n)$ in the $P3$ structure becomes similar to that in the $P1$ structure, suggesting that the pseudogap NFL state represents a more correlated NFL state. This implies that the two entangled energy bands in the $P3$ structure result in a metallic state with reduced correlations at identical $U$ and $J$ values.

We also examine the evolution of $\mathrm{Im}\,\Sigma(i\omega_n)$ during the metal-insulator transition by increasing $U$ at fixed $J = 1$ eV (Figs. S5**c,d**). As $U$ increases, the magnitude of $\mathrm{Im}\,\Sigma(i\omega_n)$ increases, leading to a transition from metallic states to insulating states for both structures. As discussed in detail in the main text, at $J = 1$ eV, both structures are metallic NFL states at $U = 3$ and $5$ eV, and become charge-transfer Mott insulating states at larger $U$. For the $P3$ structure, the metallic states maintain their bad-metallic NFL characteristics with increasing $U$, since the slope remains the same even as it becomes an insulator. For the $P1$ structure, the metallic states also maintain their NFL characteristics with increasing $U$, but the slope changes sign upon becoming an insulator. In the insulating states, both structures exhibit low-frequency values of $\mathrm{Im}\Sigma(i\omega_n)$ approaching zero at zero imaginary frequency, as there are no electrons to scatter.

### 4.2. Non-correlated orbitals

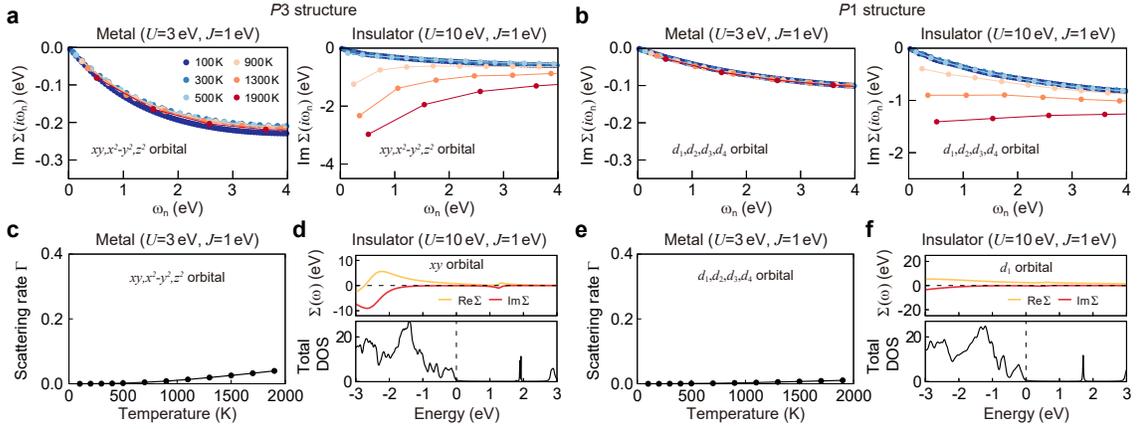

FIG. S6: **Self energy and scattering rate of non-correlated orbitals. a,b** Temperature evolution of the imaginary part of the self energy on the imaginary axis, $\text{Im}\Sigma(i\omega_n)$, for **a** the $P3$ structure and **b** the $P1$ structure. We display $\text{Im}\Sigma(i\omega_n)$ for the non-correlated orbitals by averaging them (see Fig. 4 in the main text for the self-energy of correlated orbitals). **c** Scattering rate of non-correlated orbitals in the metallic state of the $P3$ structure as a function of temperature. The scattering rate is averaged over the $d_{xy}$, $d_{x^2-y^2}$, and $d_{z^2}$ orbitals. **d** Self energy of the $d_{xy}$ orbital on the real axis and total DOS of the insulating state of the $P3$ structure. **e** Scattering rate of non-correlated orbitals in the metallic state of the $P1$ structure as a function of temperature. The scattering rate is averaged over the $d_1$, $d_2$, $d_3$, and $d_4$ orbitals. In **c,e**, the scattering rate $\Gamma \sim -\text{Im}\Sigma(i0^+)$ is obtained using a linear fit over the first two Matsubara frequencies. **f** Self energy of the $d_1$ orbital on the real axis and total DOS of the insulating state of the $P1$ structure.

For completeness, we also provide a detailed analysis on the self energy of the non-correlated orbitals in Fig. S6: $d_{xy}$, $d_{x^2-y^2}$, and $d_{z^2}$ orbitals for the $P3$ structure, and $d_1$, $d_2$, $d_3$, and $d_4$ orbitals for the $P1$ structure, which are away from the Fermi level. In the metallic states, these non-correlated orbitals are Fermi liquid states for both structures, as confirmed by the first Matsubara frequency rule in Supplementary Note 3. The low-frequency behaviour of $\text{Im}\,\Sigma(i\omega_n)$ for the non-correlated orbitals in metallic states shows linear behaviour and zero scattering rate at low temperatures (left panels in Figs. S6**a,b**). The scattering rates of the non-correlated orbitals for both structures exhibit quadratic temperature dependence (Figs. S6**c,e**), as expected for Fermi liquid states. In the insulating states for both structures, the real and imaginary parts of the self energy for the non-correlated orbitals on the real axis exhibit featureless values close to zero, without a pole structure within the gap (Figs. S6**d,f**). This confirms that the non-correlated orbitals are not relevant for Mott gap opening.

10


\fi

\end{document}